\documentclass[aps,pre,longbibliography,notitlepage,floatfix,nofootinbib,rmp,superscriptaddress]{revtex4-1}

\usepackage{bm,amsmath,amssymb,graphicx,stmaryrd,float,subcaption,upgreek,sidecap,MnSymbol,mathtools,lipsum}
\usepackage{epstopdf}
\usepackage{physics}
\usepackage[abs]{overpic}
\usepackage[justification=raggedright]{caption}
\usepackage{enumitem}
\usepackage{tikz}
\usepackage{wasysym}
\usepackage{relsize}
\usepackage{pifont}
\usepackage{xcolor}
\usepackage{tabularx}

\newcommand{\reddiamond}{\color{red} \text{\smaller[2.0] $\!\!\blacklozenge$}}
\newcommand{\bluediamond}{\color{blue} \text{\smaller[2.0] $\!\!\blacklozenge$}}
\newcommand{\bladiamond}{\color{black} \text{\smaller[2.0] $\!\!\blacklozenge$}}
\newcommand{\browndiamond}{\color[rgb]{0.64,0.16,0.16}  \text{\smaller[2.0] $\!\!\! \blacklozenge$}}

\newcommand{\redstar}{\color{red} \text{\smaller[3]$\bigstar$}}
\newcommand{\bluestar}{\color{blue} \text{\smaller[3]$\bigstar$}}

\newcommand{\redsquare}{\color{red} \text{\smaller[2.0] $\!\!\blacksquare$}}
\newcommand{\bluesquare}{\color{blue} \text{\smaller[2.0] $\!\!\blacksquare$}}
\newcommand{\blasquare}{\color{black} \text{\smaller[2.0] $\!\!\blacksquare$}}
\newcommand{\brownsquare}{\color[rgb]{0.64,0.16,0.16}  \text{\smaller[2.0] $\!\!\! \blacksquare$}}

\newcommand{\redtriangle}{\color{red} \text{\smaller[2.0] $\!\!\! \blacktriangle$}}
\newcommand{\bluetriangle}{\color{blue} \text{\smaller[2.0] $\!\!\! \blacktriangle$}}
\newcommand{\blatriangle}{\color{black} \text{\smaller[2.0] $\!\!\! \blacktriangle$}}
\newcommand{\browntriangle}{\color[rgb]{0.64,0.16,0.16}  \text{\smaller[2.0] $\!\!\! \blacktriangle$}}

\newcommand{\hollbluetriangleright}{\color{blue} \text{\smaller[1.0] $\!\!\triangleright$}}
\newcommand{\hollbluetriangleleft}{\color{blue} \text{\smaller[1.0] $\!\!\triangleleft$}}
\newcommand{\hollredtriangleright}{\color{red} \text{\smaller[1.0] $\!\!\triangleright$}}
\newcommand{\hollredtriangleleft}{\color{red} \text{\smaller[1.0] $\!\!\triangleleft$}}

\newcommand{\redtriangleright}{\color{red} \text{\larger[1.0] $\!\!\! \blacktriangleright$}}
\newcommand{\bluetriangleright}{\color{blue} \text{\larger[1.0] $\!\!\! \blacktriangleright$}}
\newcommand{\redtriangleleft}{\color{red} \text{\larger[1.0] $\!\!\! \blacktriangleleft$}}
\newcommand{\bluetriangleleft}{\color{blue} \text{\larger[1.0] $\!\!\! \blacktriangleleft$}}

\graphicspath{{images/}}

\begin{document}
	\title{Double-eigenvalue bifurcation and multistability in serpentine \\ strips with tunable buckling behaviors}
	\author{Qiyao Shi}
	\affiliation{Department of Mechanics and Aerospace Engineering, Southern University of Science and Technology, Shenzhen, China 518055}

\author{Weicheng Huang}
	\email{weicheng.huang@ncl.ac.uk}
	\affiliation{School of Engineering, Newcastle University, Newcastle upon Tyne, NE1 7RU, UK}

	\author{Tian Yu}
	\email{jhyutian@vt.edu}
	\affiliation{Department of Mechanics and Aerospace Engineering, Southern University of Science and Technology, Shenzhen, China 518055}

 \author{Mingwu Li}
	\email{limw@sustech.edu.cn}
	\affiliation{Department of Mechanics and Aerospace Engineering, Southern University of Science and Technology, Shenzhen, China 518055}
	\date{\today}

	\begin{abstract}
	Serpentine structures, composed of straight and circular strips, have garnered significant attention as potential designs for flexible electronics due to their remarkable stretchability. When subjected to stretching, these serpentine strips buckle out of plane, and previous studies have identified two distinct buckling modes whose order of appearance may interchange in serpentine structures with a single cell. In this study, we employ anisotropic rod theory to model serpentine strips as a multi-segment boundary value problem (BVP), with continuity conditions enforced at the interface between the straight and curved strips. We solve the BVP using methods of continuation, and our results reveal that: 1) the exchange of the two buckling modes in a single-cell serpentine strip is induced by a double-eigenvalue and associated secondary bifurcations, which also alter the stability of the two buckling modes; 2) a variety of stable states with reversible symmetry can be manually obtained in tabletop models and are found to be disconnected from the planar branch in numerical continuation. Furthermore, we demonstrate that modulating the strip thickness across different cells leads to the initiation of buckling in the thinnest section, thereby allowing for the tuning of buckling modes in serpentine strips. In structures with two cells, the sequence of the two buckling modes can also be controlled by designing serpentine strips with nonuniform height. This work could enhance the mechanical design of serpentine-interconnect-based flexible structures and could have applications in multistable actuators and mechanical memory devices.   
	\end{abstract}

	\maketitle

\section{Introduction}	
In recent years, there have been a surging interest in the study of the mechanics of slender structures due to their potential applications in designing deployable and morphing structures \cite{panetta2019x,yu2021numerical,liu2023deployable,pillwein2020elastic,celli2020compliant,yu2023continuous,yang2023morphing}, complex 3D shapes \cite{baek2020smooth,liu2019postbuckling,liu2020tapered, huang2024integration}, flexible electronics and actuators \cite{zhang2013buckling,cheng2021anti,taghavi2018electro}, flexbile medical device \cite{till2017elastic,wang2021evolutionary}, and soft robots \cite{chi2022snapping,huang2020dynamic,huang2023modeling}. 
These application scenarios are often achieved through large deformations, bistability/multistability, and snapping behaviors of slender structures such as strips and rods. For example, two strips coupled at their ends form a hair-clip-like bistable structure called ``bigon'', which serves as a building block for constructing deployable and multistable structures, such as a ``bigon ring'' \cite{yu2021numerical}. The snapping mechanism in the hair clip has been used to design a high-speed flexible swimmer \cite{chi2022snapping} and a swimming robot \cite{chen2018harnessing}. Gridshells coupling multiple rods create complicated 3D surfaces involving large deformations. Recently, X-shell consisting of strips with specific geometric design have been shown to morph from flat to 3D shapes \cite{panetta2019x}. Slender structures could be folded into multiply covered loops \cite{starostin2022forceless,audoly2015buckling}, which is utilized in bandsaw blades for compact packaging and is further realized in complex thin structures through careful geometric and mechanical design \cite{mouthuy2012overcurvature,yu2021numerical,yu2023continuous,lu2023multiplei,lu2023multipleii,wu2021ring}.

The aforementioned applications often requires an understanding of the mechanics of thin structures and, sometimes, an inverse design to achieve specific mechanical and geometric behaviors. 
Slender structures such as thin strips and rods can easily undergo large deformations, exhibit buckling behaviors, and have multiple equilibria, resulting in rich and complex nonlinear mechanical behaviors. While a single strip or rod can exhibit a variety of bifurcations, multistability, and complicated buckling patterns \cite{huang2020shear,yu2019bifurcations,domokos2005multiple,goss2005experiments,korte2011triangular,audoly2015buckling,lazarus2013continuation,moulton2018stable,sano2019twist,huang2024exploiting}, coupling multiple strips or rods with joints through tailored geometric designs can create deployable structures \cite{lu2023multiplei,lu2023multipleii,panetta2019x}, gridshells \cite{baek2018form}, and complex surfaces \cite{baek2020smooth,ren20213d}. 
Large deformations of thin strips and rods are generally dominated by bending and twisting, with negligible stretching of their centerlines. The Kirchhoff rod model and inextensible strip theory have been shown to accurately capture the nonlinear mechanics of thin rods and strips. Discrete elastic rods and strips have been developed to efficiently model the mechanical behaviors of these slender structures \cite{bergou2008discrete, korner2021simple}.

Recently, structures that concatenate multiple identical serpentine cells, each composed of straight and circular strips, have been proposed for the design of flexible electronics \cite{zhang2013buckling,zhang2014hierarchical}. These designs minimize axial strains in the material because the deformation is dominated by bending and twisting of the strips. Combined with a flexible elastomer substrate, serpentine interconnects can achieve remarkable stretchability without causing significant plastic deformations to the materials \cite{bian2017buckling}. Serpentine interconnects have been used to design flexible supercapacitors \cite{rajendran2019all} and stretchable batteries \cite{xu2013stretchable}.
The mechanics of serpentine interconnects have been extensively studied. Upon stretching, serpentine interconnects could remain in-plane as a planar structure or experience lateral buckling, deforming into a 3D shape. This behavior depends on the competition between in-plane bending stiffness and out-of-plane bending and twisting stiffness \cite{zhang2013mechanics,wang2023substantial,zhang2013buckling}. The nonlinear mechanics of non-buckling serpentines that remain in-plane up to large deformations have been investigated through curved beam theory, finite element modeling, and plane elasticity \cite{yang2017elasticity,liu2019experimentally}. When the width of the serpentine interconnect is much larger than its out-of-plane thickness, the in-plane bending stiffness is significantly greater than the out of plane bending and twisting stiffness, causing the serpentine strip to buckle out of plane, with two different buckling modes observed \cite{zhang2013buckling}. 
For serpentine strips with a single cell, either of the two buckling modes may occur first, depending on the dimensionless height of the structure \cite{zhang2013buckling}. However, the cause of the change in the buckling sequence has not been fully addressed. 

In the current paper, we focus on the nonlinear mechanics of serpentine strips where the in-plane width being much larger than the thickness and significantly smaller than the length. This configuration causes the strip to buckle out of plane immediately when subjected to stretching. We aim to address the following questions: \emph{What causes the exchange of the buckling modes in serpentine strips, as observed in the original study \cite{zhang2013buckling}? Can the buckling pattern be tuned by adjusting the geometry of the structure? } 
We use anisotropic rod model to study the large deformations of serpentine strips, formulating it as a boundary value problem (BVP) which is solved through numerical continuation~\cite{ahsan2022methods}. Rods and strips with clamping boundary conditions are known to exhibit reversible symmetry, and the solutions either possess this symmetry themselves or appear as reversible pairs \cite{starostin2022forceless, van1998lock, champneys1996multiplicity, domokos2001hidden}. We leverage the symmetry properties of the rod equations to identify symmetric pairs in serpentine strips. Numerical continuation of the anisotropic rod equations allows us to study bifurcations and trace the equilibrium branches in serpentine strips, revealing the underlying mechanisms that cause the exchange of the two buckling modes. These continuation results are further confirmed by experiments.

The rest of the paper is organized as following. In section \ref{sec:experiments}, we introduce the experimental setup for investigating the buckling and multistable behaviors of serpentine strips with clamping boundary conditions. Section \ref{se:anisotropicrod} presents the anisotropic rod theory and its application to serpentine strips with reversible symmetry. In Section \ref{se:stablestates}, we compare the stable states of serpentine strips obtained from tabletop models with the numerical solutions of anisotropic rod model.
Section \ref{bifurcationdiagrams} provides bifurcation diagrams of serpentine strips and the evolution of various bifurcations/folds in a phase diagram. In Section \ref{se:tunablebuckling}, we demonstrate that by designing the geometry of serpentine strips with nonuniform thickness and height, both the buckling pattern and the buckling sequence can be tuned. Finally, Section \ref{se:discussion} includes a discussion of the current results and outlines directions for future work.

\section{Geometry of serpentine strips and the experimental set up} \label{sec:experiments}
\begin{figure}[h!]
	\centering
\includegraphics[width=0.85\textwidth]{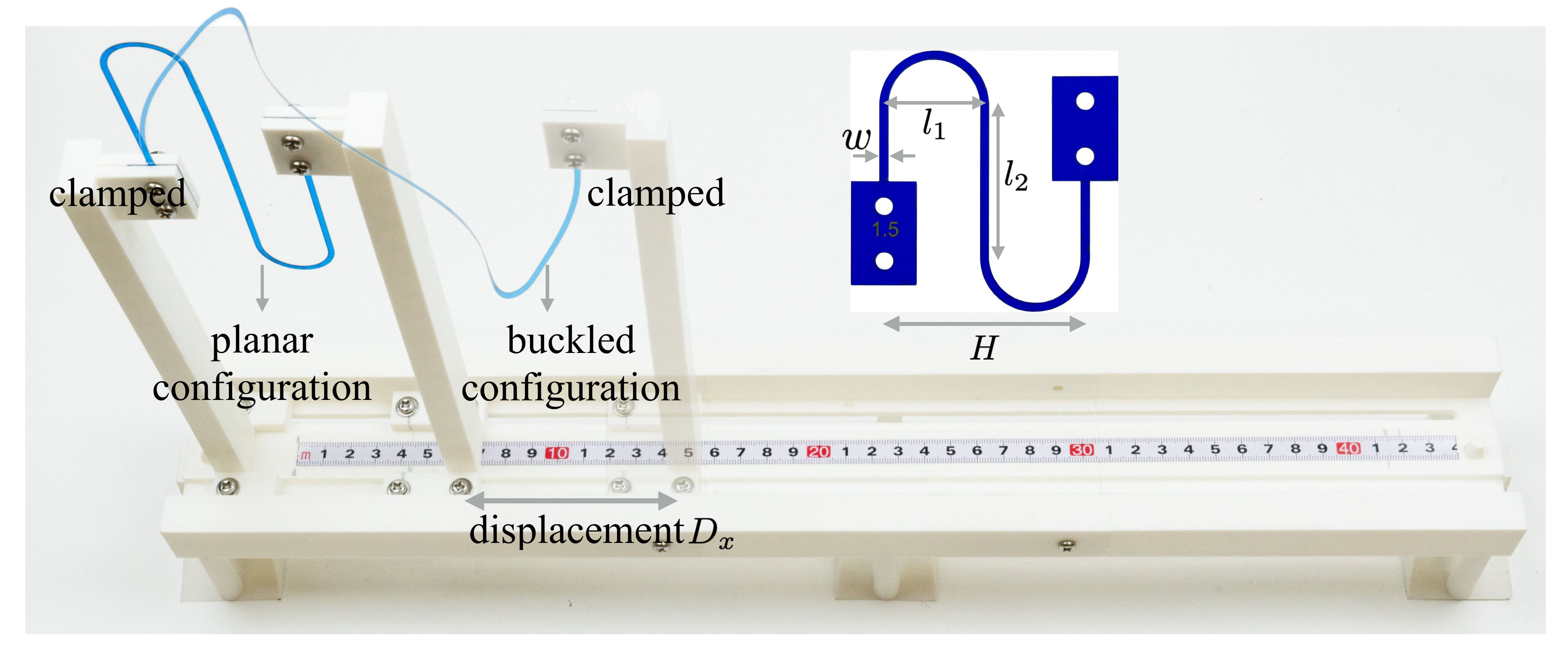}
	\caption{The geometry of a single-cell serpentine strip and the 3D-printed experimental setup. The two ends of the serpentine strip are clamped by two vertical arms, which can slide to adjust the distance between them. }\label{fig:exptsetup}
\end{figure}

A unit cell of a serpentine strip consists of three straight and two circular segments, forming a planar structure in its rest configuration (see Figure \ref{fig:exptsetup}). The central straight segment has a length $l_2$, and the two side straight segments each have a length $0.5 l_2$. The diameter of the two semicircles is $l_1$. We introduce a dimensionless height $\alpha=l_2/l_1$. 
The thickness of the strip $t$ (which is much smaller than the width $w$) is perpendicular to the structural plane, making the in-plane bending stiffness of the strip significantly larger than the out of plane bending and twisting stiffness. Multiple unit cells can be concatenated to form a larger serpentine strip.

We have fabricated an experimental device using 3D printing to clamp and stretch serpentine strips. The experimental setup consists of a base with a groove and two sliding arms that secure the ends of the strip. 
A flexible ruler is affixed to the base to measure the displacement applied to the sliding end. 
The serpentine strips, laser-cut from PVC sheets, have a thickness of $t=0.24$ mm and a width of $w=2.4$ mm ($w/t=10$). The radius of the semicircle is set to $5w$ (i.e. $l_1=10w$) to ensure the structure remains slender. For wide serpentine strips with $l_1<5w$, a 2D elastic theory is more appropriate for capturing its deformations  \cite{wang2023substantial,yang2017elasticity}. Each end of the serpentine strip is connected to a rectangular panel with two holes, which are clamped tightly by the sliding arms using bolts that pass through the holes. 
The extended panels are cut together with the serpentine strip as an integrated piece.

Due to the large aspect ratio of the cross section of the serpentine strip, its in-plane bending stiffness $Etw^3/12$, is significantly greater than its out-of-plane bending stiffness $Ewt^3/12$, and twisting stiffness ($\sim Gwt^3/12$), where $E$ and $G$ correspond to the Young's modulus and shear modulus of the material, respectively. When a displacement is applied to one end, the serpentine strip smoothly buckles out of plane, and the buckled configuration remains stable up to large displacements (see the buckled configuration in Figure \ref{fig:exptsetup}). With a fixed displacement $D_x$, multiple stable states can be found by manually deforming the structure to reach different equilibria. The number of stable states depends on the dimensionless height $\alpha$ and the number of cells in the serpentine strip.

\section{Anisotropic rod model} \label{se:anisotropicrod} 

\begin{figure}[h!]
	\centering
	\includegraphics[width=0.85\textwidth]{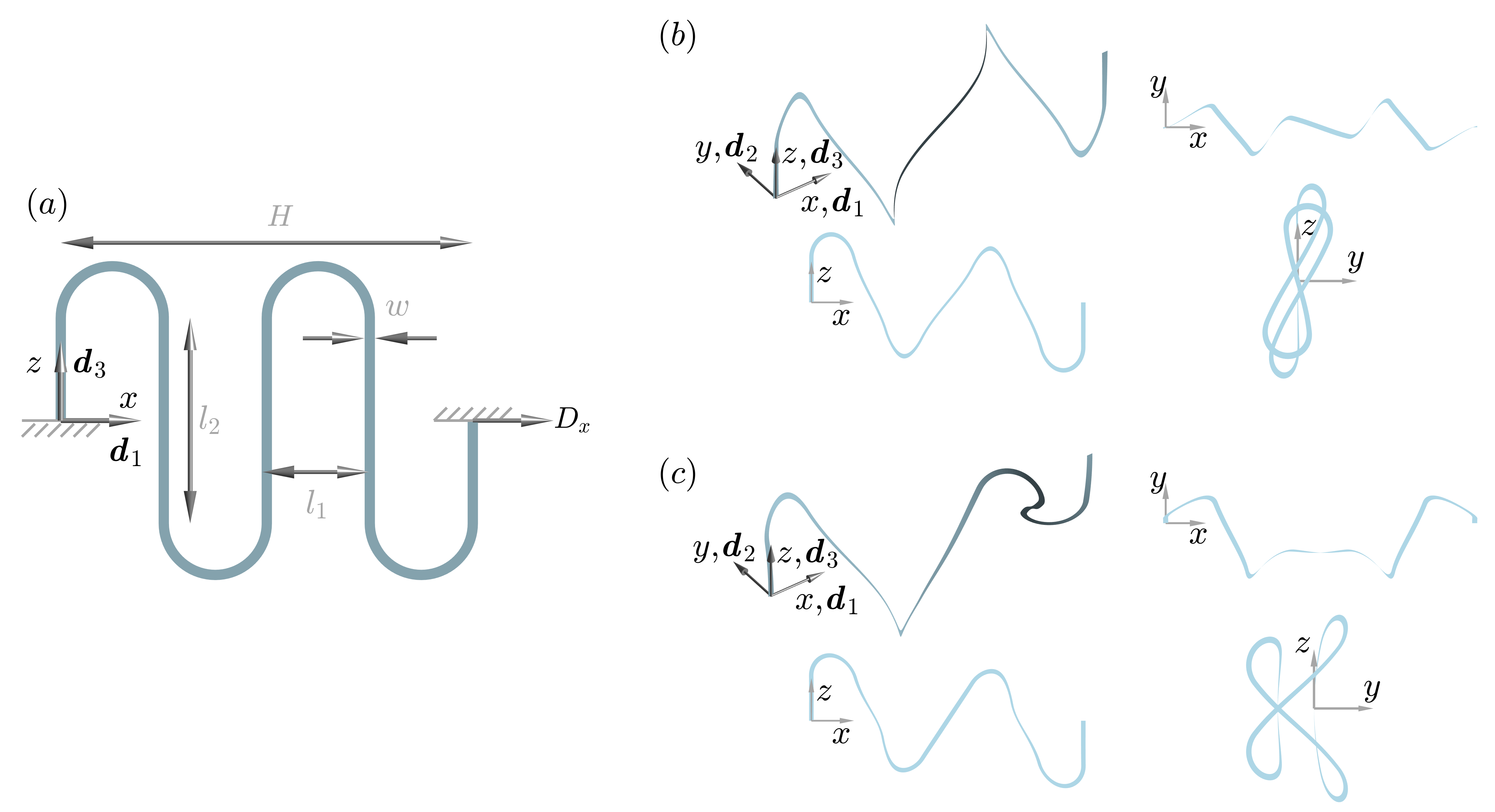}
	\caption{Out-of-plane buckling behavior of a serpentine strip. (a) Geometry of a serpentine strip with two cells ($n_c=2$). Two different buckling modes in 3D view and 2D projections: (b) $S$ mode, and (c) $M$ mode. }\label{fig:serpentinegeometry}
\end{figure}

The anisotropic rod model has been demonstrated to accurately capture the large deformation and complex bifurcation behaviors of narrow strips with $w << l$ and $w \sim O(10t)$ \cite{moulton2018stable,yu2021numerical,yu2019bifurcations,riccobelli2021rods}.
In this study, we employ anisotropic rod theory to study the nonlinear mechanics of serpentine strips with a fixed $w/t$ ratio of 10, unless otherwise stated.
Figure \ref{fig:serpentinegeometry} illustrates a serpentine strip with two cells. The left end of the structure is fixed at the origin of a Cartesian coordinate system $x-y-z$, while the right end is subjected to displacement $D_x$ along the $x$ axis. An orthonormal right-handed material frame $( \bm{d}_1, \bm{d}_2, \bm{d}_3 )$ is attached to the centerline of the strip, with $\bm{d}_1 $, $\bm{d}_2$ and $ \bm{d}_3$ aligned with the width of the strip, the thickness of the strip, and the local tangent of the centerline, respectively. At the left clamping end, $\bm{d}_1 $, $\bm{d}_2$ and $ \bm{d}_3 $ are aligned to $x$, $y$, and $z$ direction, respectively. The dimensionless geometric parameters include $l_2/l_1$, the cross-sectional aspect ratio of the ribbon $w/t$, and the number of unit cells $n_c$. The horizontal span $H$ of the system is given by $2n_c l_1$. 
Figures \ref{fig:serpentinegeometry}b and \ref{fig:serpentinegeometry}c illustrate the two buckling modes identified in the original study \cite{zhang2013buckling}, whose projections onto the $x-y$ plane are antisymmetric and symmetric about the midpoint, respectively. However, this antisymmetry and symmetry do not apply to the 3D configurations, which exhibit reversible symmetry, as discussed later. Instead of referring to them as the antisymmetric and symmetric modes, as used in the original study \cite{zhang2013buckling}, we designate the buckling modes shown in Figures \ref{fig:serpentinegeometry}b and \ref{fig:serpentinegeometry}c as the $S$ and $M$ modes, respectively. 
Additionally, we will present other stable states that can be manually achieved by deforming the structure. 
Based on anisotropic rod model, the force and moment equilibrium equations can be summarized as, 
\begin{equation}\label{eq:F&Mbalance}
\begin{aligned}
\bm{N}' &=\bm{0} \,,  \\
\bm{M}' +\bm{d_3} \times \bm{N} &=\bm{0} \,, \\
\end{aligned}
\end{equation}
where a prime denotes a derivative with respect to the arc length $s$ along the centerline ($s \in [0,L]$), and $L$ represents the total length of the serpentine strip. $\bm{N}$ and $\bm{M}$ denote the contact forces and moments, respectively. The motion of the material frame is described by $\bm{d}_i'=\bm{\omega} \times \bm{d}_i$, with $\bm{\omega}\!=\!\kappa_{1} \bm{d}_1 + \kappa_{2} \bm{d}_2 +\tau \bm{d}_3$, where $\kappa_{1}, \kappa_{2}$ and $\tau$ represent the two bending curvatures and the twist of the centerline, respectively.
The contact forces and moments can be resolved in the material frame as $\bm{N}=N_1 \bm{d_1} +N_2 \bm{d_2} +N_3 \bm{d_3}$ and $\bm{M}=M_1 \bm{d_1} +M_2 \bm{d_2} +M_3 \bm{d_3}$.

We assume linear constitutive relations $M_1=EI_1 \kappa_1 $, $M_2=EI_2 (\kappa_2 - \kappa_{20})$, and $M_3= GJ \tau$, where $E$ and $G$ are the Young's modulus and shear modulus, respectively; $EI_1$, $EI_2$, and $GJ$ represent the two bending rigidities and the torsional rigidity, respectively. The term $\kappa_{20}$ denotes the in-plane rest curvature, which is constant for the semicircle and zero for the straight segment. By substituting the resolved forces/moments along with the kinematic relationship $\bm{d}_i'=\bm{\omega} \times \bm{d}_i$ into Equation \eqref{eq:F&Mbalance} and defining the stiffness ratios $a=EI_1/(GJ)$ and $b=EI_2/(GJ)$, we obtain
\begin{equation}\label{eq:F&Mequilibrium} 
\begin{aligned}
& N_1'  = N_2 \tau -N_3 \kappa_2 \, , N_2'  = -N_1 \tau + N_3 \kappa_1  \, , N_3'  =  -N_2 \kappa_1 + N_1 \kappa_2 \, , \\
& a \kappa_1' = b (\kappa_2 - \kappa_{20}) \tau - \tau \kappa_2 + N_2 \, , b \kappa_2 ' = -a \kappa_1  \tau + \tau \kappa_1  - N_1 \, , \tau'  = -b (\kappa_2 - \kappa_{20} )  \kappa_1 + a \kappa_1 \kappa_2   \, .
\end{aligned}
\end{equation}

Slender rods and strips subjected to clamping boundary conditions are known to exhibit reversible symmetry \cite{starostin2022forceless, van1998lock,champneys1996multiplicity,domokos2001hidden}. In other words, the equilibrium configuration either possesses reversibility itself or appear as a reversible pair. 
The equilibrium Equation \eqref{eq:F&Mequilibrium} is invariant under the following transformations,

\begin{equation}\label{eq:reversiblesymmetry} 
\begin{aligned} 
& R_1: s \rightarrow L-s , \, \kappa_{20} \rightarrow -\kappa_{20}, \, \kappa_1 \rightarrow -\kappa_1, \, \kappa_2 \rightarrow -\kappa_2, \, \kappa_3 \rightarrow -\kappa_3, \, N_1 \rightarrow N_1, \, N_2 \rightarrow N_2, \, N_3 \rightarrow N_3.  \\
& R_2: s \rightarrow L-s , \, \kappa_{20} \rightarrow -\kappa_{20}, \, \kappa_1 \rightarrow \kappa_1, \, \kappa_2 \rightarrow -\kappa_2, \, \kappa_3 \rightarrow \kappa_3, \, N_1 \rightarrow N_1, \, N_2 \rightarrow -N_2, \, N_3 \rightarrow N_3. \\
& K: s \rightarrow s , \,\,\,\,\,\,\,\,\,\,\,\, \kappa_{20} \rightarrow \kappa_{20}, \, \kappa_1 \rightarrow -\kappa_1, \, \kappa_2 \rightarrow \kappa_2, \, \kappa_3 \rightarrow -\kappa_3, \, N_1 \rightarrow N_1, \, N_2 \rightarrow -N_2, \, N_3 \rightarrow N_3,
\end{aligned}
\end{equation}
where $R_1$ and $R_2$ represent reversing involutions, and $K$ corresponds to a non-reversing involution. Note that $\kappa_{20}$ changes sign for $R_1$ and $R_2$ but remains the same for $K$.

In serpentine strips, the rest in-plane curvature $\kappa_{20}$ is discontinuous at the interfaces between straight and circular segments, and this discontinuity extends to the deformed curvature $\kappa_2$. We model serpentine strips as a multiple-point boundary value problem and impose the continuity of forces, moments, and material frames at the interfaces as boundary conditions. Specifically, each serpentine cell is divided into five segments (two circular and three straight segments), with a full set of governing equations assigned to each segment. These segments are coupled through boundary conditions at the interfaces. Utilizing the technique described in \cite{ascher1981reformulation}, we reformulate the multi-segment boundary value problem as a standard two-point boundary value problem, which is solved by conducting numerical continuation with AUTO 07P \cite{doedel2007auto}. Details of this process are included in Supplementary Material \ref{appse:MPbvpToTPBVP}.
To determine the stability of a numerical solution, we employ a geometric mechanics method \cite{borum2020helix,borum2020infinitely} and test for the existence of a conjugate point within the integral interval $[0 \,, L]$. Additional details can be found in Supplementary Material \ref{appse:stabilitytest}.

\section{Stable states in experiments and numerical solutions} \label{se:stablestates} 

\begin{figure}[h!]
	\centering
\includegraphics[width=0.95\textwidth]{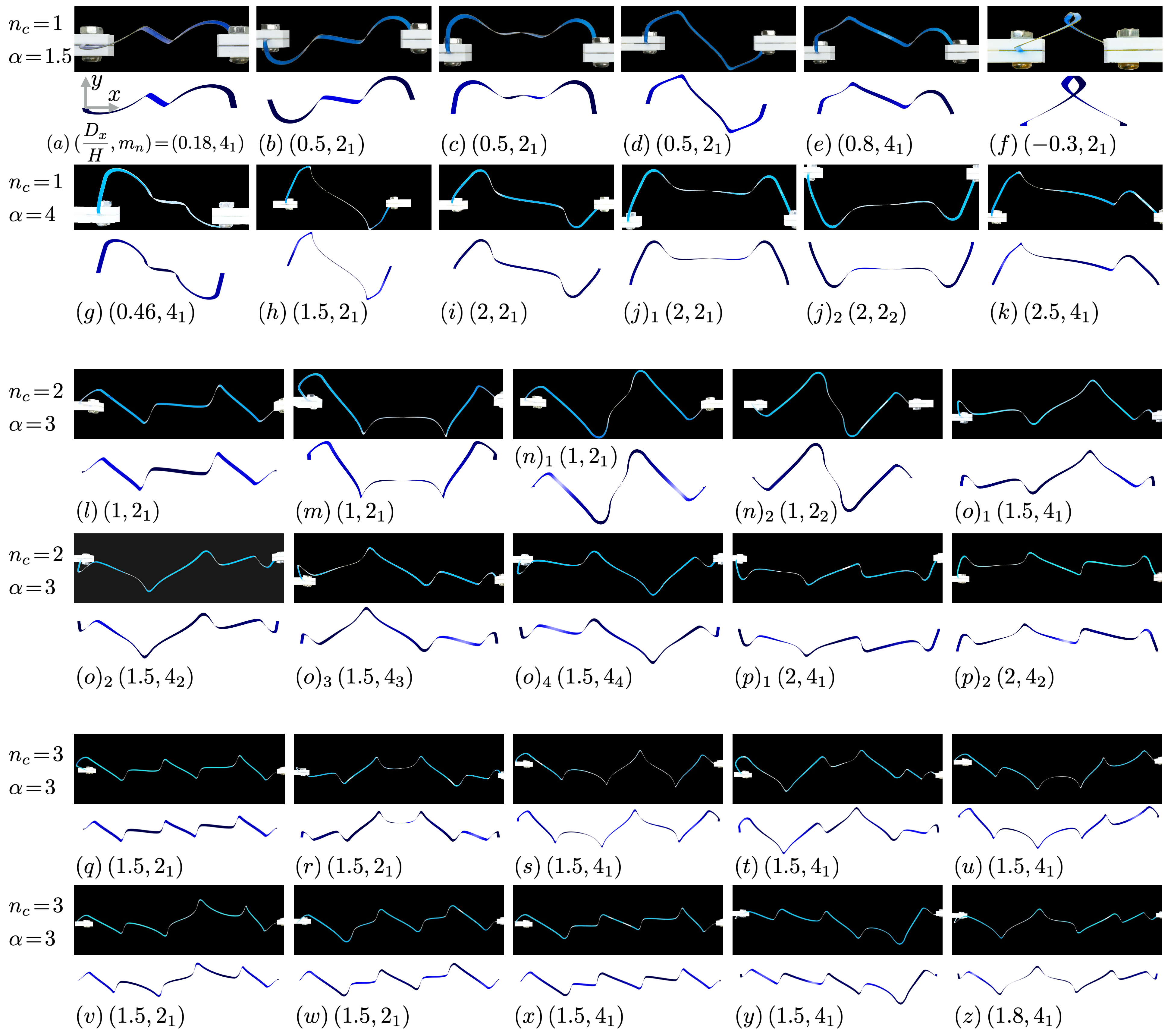}
	\caption{2D $x-y$ projection of stable equilibria in serpentine strips with different number of cells ($n_c$), dimensionless heights ($\alpha$), and stretches ($D_x/H$). Configurations with black and white backgrounds represent tabletop models and the numerical solutions of the anisotropic rod model, respectively. (a-f) One cell with $\alpha=1.5$. (g-k) One cell with $\alpha=4$. (l-p) Two cells with $\alpha=3$. (q-z) Three cells with $\alpha=3$. Here, $m_n$ in each subfigure represents the $n_\text{th}$ state in a group of $m$ symmetric shapes.}\label{fig:expts_simulations}
\end{figure}

In this section, we present the stable states of serpentine strips with different number of cells. First, we identify the stable states through tabletop models by manually deforming the structure to reach various equilibrium. Then, we find the corresponding states through numerical continuation of the anisotropic rod model. Although we are able to find a variety of stable states in serpentine strips, our method does not identify all possible stable states, nor is that our aim in this study.
Figure \ref{fig:expts_simulations} displays the $x-y$ projection of the stable states with different dimensionless heights $\alpha$ and stretches $D_x/H$. In each subfigure, $m_n$ represents the $n_\text{th}$ state in a group of $m$ symmetric shapes.

Due to symmetry, the buckled equilibria always appear in pairs. Out-of-plane buckling selects a minimum of two shapes that are symmetric about the $x-z$ plane. 
If a stable state exhibits self-reversibility about its midpoint, its mirror image about the $x-z$ plane (corresponding to the $K$ operation in Equation \eqref{eq:reversiblesymmetry}) forms a second state, resulting in a pair of symmetric states. In this scenario, application of the reversible operations $R_1$ and $R_2$ either produces the shape itself or its mirror image, thus not creating new states. However, a stable shape lacking self-reversibility can generate three additional shapes through the $R_1$, $R_2$ and $K$ operations, resulting in a group of four symmetric shapes.
For example, the configurations in Figures 3b-3d, 3f, and 3h-3j exhibit self-reversibility and each represents one shape in a symmetric pair. Figures \ref{fig:expts_simulations}j$_1$ and \ref{fig:expts_simulations}j$_2$ display both configurations in a symmetric pair. The states in Figures \ref{fig:expts_simulations}a, \ref{fig:expts_simulations}e, \ref{fig:expts_simulations}g and \ref{fig:expts_simulations}k do not exhibit self-reversibility, with each representing one shape in a group of four symmetric states. Figures \ref{fig:expts_simulations}o$_1$, \ref{fig:expts_simulations}o$_2$, \ref{fig:expts_simulations}o$_3$ and \ref{fig:expts_simulations}o$_4$ present all the shapes in a group of four symmetric states.   
We achieve good agreement between tabletop models (configurations with black background) and numerical solutions of the anisotropic rod model (renderings with white background), demonstrating the accuracy of the anisotropic rod model in capturing the nonlinear mechanics of serpentine strips. 
Compressing serpentine strips also results in out-of-plane buckling behavior. This study primarily focuses on stretching-induced buckling but includes one example of a symmetric pair with compressive displacement (Figure \ref{fig:expts_simulations}f).  

As the number of cells increases, the number of stable states also increases. Figures \ref{fig:expts_simulations}l-\ref{fig:expts_simulations}p present stable shapes for $(n_c,\alpha)=(2,3)$. Figures \ref{fig:expts_simulations}l and \ref{fig:expts_simulations}m each display one shape in a symmetric pair. Figures \ref{fig:expts_simulations}n$_1$ and \ref{fig:expts_simulations}n$_2$ present both configurations in a symmetric pair. Figures \ref{fig:expts_simulations}p$_1$ and \ref{fig:expts_simulations}p$_2$ show two configurations in a group of four symmetric states, while Figures \ref{fig:expts_simulations}o$_1$- \ref{fig:expts_simulations}o$_4$ depict all the states in a group of four symmetric shapes. In total, we identify 14 stable shapes for $(n_c,\alpha)=(2,3)$.  
Figures \ref{fig:expts_simulations}q-\ref{fig:expts_simulations}z present stable states of serpentine strips with $(n_c,\alpha)=(3,3)$. Figures \ref{fig:expts_simulations}q, \ref{fig:expts_simulations}r, \ref{fig:expts_simulations}v and \ref{fig:expts_simulations}w each display one shape in a symmetric pair. The remaining figures each present one state in a group of four symmetric shapes. In total, we identify 32 stable equilibria for $(n_c,\alpha)=(3,3)$.

The original work \cite{zhang2013buckling} found that the first buckling mode of a serpentine strip with a single cell ($n_c=1$) could change from the $S$ mode to the $M$ mode as the dimensionless height $\alpha$ increases. However, serpentine strips with two or more cells (i.e. $n_c \ge 2$) consistently buckle into the $S$ mode, regardless of the value of $\alpha$. The cause of this exchange in buckling modes remains unknown.
This behavior aligns with our experimental observations. For instance, with $(n_c,\alpha)=(1,1.5)$, the structure first buckles into the $S$ mode (Figure \ref{fig:expts_simulations}b), and manual deformation is required to achieve the $M$ mode (Figure \ref{fig:expts_simulations}c). Conversely, for $\alpha=4$, the structure initially buckles into the $M$ mode (Figures \ref{fig:expts_simulations}j$_1$ and \ref{fig:expts_simulations}j$_2$), and manual deformation is needed to achieve the $S$ mode (Figure \ref{fig:expts_simulations}h). 
Additionally, serpentine strips with $n_c=2$ and $n_c=3$ consistently buckle into the $S$ mode (Figure \ref{fig:expts_simulations}l and \ref{fig:expts_simulations}q), regardless of the value of $\alpha$, with manual deformation required to reach the $M$ mode (Figure \ref{fig:expts_simulations}m and \ref{fig:expts_simulations}r). 
In the following section, we present the bifurcation and phase diagrams associated with the buckling behaviors of serpentine strips, which illustrate the evolution of various states.

\section{Bifurcation and Phase diagrams} \label{bifurcationdiagrams}

Bifurcation diagrams are obtained through numerical continuation of the boundary value problem (based on anisotropic rod model) using AUTO 07P \cite{doedel2007auto}, with the dimensionless stretch $D_x/H$ as the bifurcation parameter. For the vertical axis of the bifurcation diagram, we select the out-of-plane displacement $y_0$ at the straight-circular interface closest to the midpoint and having $z>0$. This interface is marked by a red disk in an inset embedded within the bifurcation diagram. In all bifurcation diagrams presented in this study, black and grey curves represent stable and unstable equilibria, respectively. Black and grey disks denote bifurcation and fold points, respectively. We also display the 3D shape and corresponding $x-y$ projection of some solutions marked by polygonal symbols in the bifurcation diagram. 
The solution curves are presented up to a reasonable value of $D_x/H$, avoiding reaching the inextensible limit of the strip.
This study focuses on stable states, and some unstable branches are truncated with a grey cross. We do not aim to identify all stable states but highlight the exchange of buckling modes and some stable states that are not identified in the original study \cite{zhang2013buckling}.

\begin{figure}[h!]
	\centering
	\includegraphics[width=0.95\textwidth]{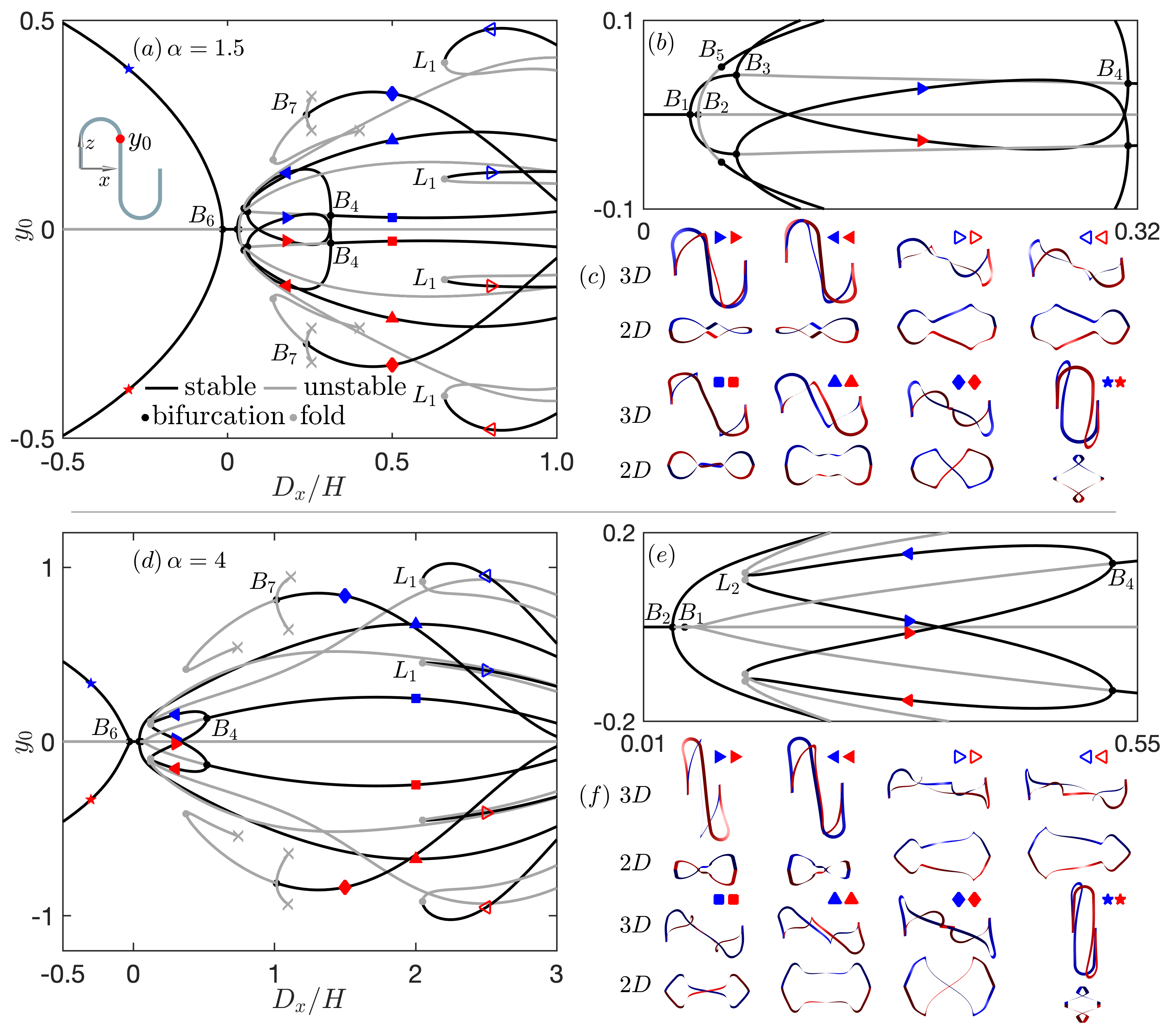}
	\caption{Bifurcation diagrams of serpentine strips with a single cell and different dimensionless heights $\alpha$. (a) $\alpha=1.5$. (b) An inset of (a). (c) Renderings corresponding to stable equilibria in (a). (d) $\alpha=4$. (e) An inset of (d). (f) Renderings corresponding to stable equilibria in (d).}\label{fig:SerpentineNc1Aniso10}
\end{figure}

Figure \ref{fig:SerpentineNc1Aniso10} presents numerical results for serpentine strips with $(n_c,w/t)$ fixed to $(1,10)$.
Figures \ref{fig:SerpentineNc1Aniso10}a-\ref{fig:SerpentineNc1Aniso10}c and \ref{fig:SerpentineNc1Aniso10}d-\ref{fig:SerpentineNc1Aniso10}f show the bifurcation diagrams and stable states for $\alpha=1.5$ and $\alpha=4$, respectively. The solution renderings in Figures \ref{fig:SerpentineNc1Aniso10}c and \ref{fig:SerpentineNc1Aniso10}f represent the equilibria marked by the polygonal symbols in Figures \ref{fig:SerpentineNc1Aniso10}a and \ref{fig:SerpentineNc1Aniso10}d, respectively. These stable configurations and their 2D projections onto the $x-y$ plane are presented as pairs of shapes (blue and red) that are symmetric about the $x-z$ plane.
Figures \ref{fig:SerpentineNc1Aniso10}b and \ref{fig:SerpentineNc1Aniso10}e provide detailed views of the various bifurcations shown in Figures \ref{fig:SerpentineNc1Aniso10}a and \ref{fig:SerpentineNc1Aniso10}d, respectively.   

Starting from the rest configuration with $D_x/H=0$, increasing the stretch causes the serpentine strip to buckle out of plane, destabilizing the planar branch (i.e., the line with $y_0=0$). 
For $\alpha \!=\! 1.5$ in Figure \ref{fig:SerpentineNc1Aniso10}a, the serpentine strip first buckles into a pair of stable $S$ mode through a supercritical pitchfork $B_1$ (details are included in Figure \ref{fig:SerpentineNc1Aniso10}b), which soon loses stability through the supercritical pitchfork $B_3$ and then regains stability through another supercritical pitchfork $B_4$. The pair of $S$ mode ($\bluesquare$/$\redsquare$) remains stable up to large stretches, both in experiments and numerical continuation, and is unstable only within the interval bounded by $B_3$ and $B_4$. The branches bifurcated from $B_3$ and $B_4$ connect to each other and lose the self-reversibility of the $S$ mode (marked by $\redtriangleleft$/$\redtriangleright$/$\bluetriangleleft$/$\bluetriangleright$). The $\bluetriangleleft$ branch corresponds to the experimental configuration in Figure \ref{fig:expts_simulations}a. 
Shortly after $B_1$, a second primary bifurcation $B_2$ emerges and bifurcates to a pair of unstable states that are later stabilized by a subcritical pitchfork bifurcation $B_5$ (the unstable bifurcated branches are not included for clarity), which connects to the stable $M$ mode $\redtriangle$/$\bluetriangle$. The $\bluetriangle$ branch matches with the experimental configuration in Figure \ref{fig:expts_simulations}c). In summary, for $\alpha=1.5$, the buckling of serpentine strip follows $B_1$ to reach one of the $S$ modes, then follows the bifurcated branch connecting $B_3$ and $B_4$, and finally returns to the $S$ mode ($\bluesquare$/$\redsquare$) through $B_4$. 

There are additional stable branches that are disconnected from the planar branch. Several such stable branches are included in Figure \ref{fig:SerpentineNc1Aniso10}a. The $\reddiamond$/$\bluediamond$ pair, which exhibits self-reversibility, is stabilized by a subcritical pitchfork bifurcation $B_7$. The $\bluediamond$ branch of this pair corresponds to the experimental configuration in Figure \ref{fig:expts_simulations}d. Additionally, $\hollredtriangleleft$/$\hollredtriangleright$/$\hollbluetriangleleft$/$\hollbluetriangleright$ represent a group of four symmetric shapes without self-reversibility, whose stability is bounded by the fold point $L_1$. The $\hollbluetriangleleft$ branch corresponds to the experimental state in Figure \ref{fig:expts_simulations}e.  
Under a compressive displacement (i.e.  $D_x/H<0$), the serpentine strip buckles out of plane through a supercritical pitchfork bifurcation $B_6$, bifurcating to a pair of stable branches $\redstar$/$\bluestar$, with $\bluestar$ corresponding to the experimental state shown in Figure \ref{fig:expts_simulations}f.

For $\alpha=4$ in Figures \ref{fig:SerpentineNc1Aniso10}d-\ref{fig:SerpentineNc1Aniso10}f, $B_1$ and $B_2$ switch positions, and the planar branch first buckles into a pair of $M$ modes ($\redtriangle$/$\bluetriangle$) through $B_2$, which remains stable up to large stretches. Shortly after $B_2$, $B_1$ emerges and bifurcates into a pair of unstable shape that are later stabilized by the supercritical pitchfork $B_4$, which connects to the stable $S$ mode. A group of four stable shape without self-reversibility ($\redtriangleleft$/$\redtriangleright$/$\bluetriangleleft$/$\bluetriangleright$) bifurcate from two identical $B_4$ (symmetric about $y_0=0$) and lose their stability through the fold point $L_2$. In summary, for $\alpha=4$, the serpentine strip follows $B_2$ to reach one of the $M$ modes, remaining stable up to large stretch. The $S$ mode is disconnected from the planar branch, requiring manual perturbations to reach it. 
Similar to $\alpha=1.5$, other stable states disconnected to the planar branch exist. In Figure \ref{fig:SerpentineNc1Aniso10}d, we include a pair of stable branches with self-reversibility ($\reddiamond$/$\bluediamond$) and a group of four symmetric shapes without self-reversibility ($\hollredtriangleleft$/$\hollredtriangleright$/$\hollbluetriangleleft$/$\hollbluetriangleright$). Their stability is bounded by the subcritical pitchfork $B_7$ and the fold point $L_1$, respectively.

\begin{figure}[h!]
	\centering
	\includegraphics[width=0.95\textwidth]{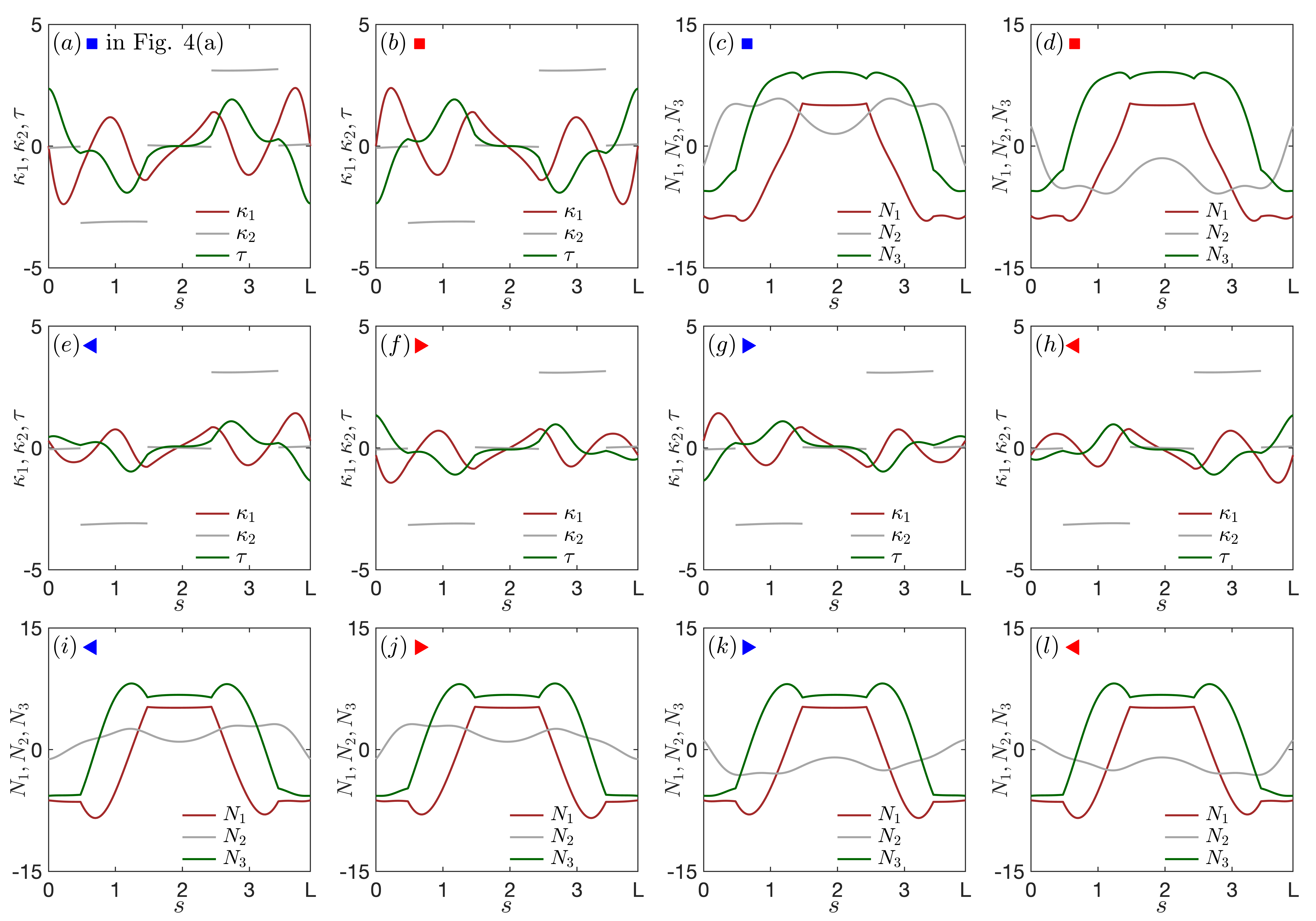}
	\caption{Curvatures ($\kappa_1$, $\kappa_2$, and $\tau$) and contact forces ($N_1$, $N_2$, and $N_3$) of stable equilibria corresponding to the solutions in Figure \ref{fig:SerpentineNc1Aniso10}a. (a-d) A symmetric pair $\bluesquare$/$\redsquare$. (e-l) A group of four symmetric shapes $\redtriangleleft$/$\redtriangleright$/$\bluetriangleleft$/$\bluetriangleright$.}\label{fig:Curvature_force}
\end{figure}

Figure \ref{fig:Curvature_force} displays the curvatures and contact forces of a pair of symmetric solutions ($\redsquare$/$\bluesquare$: Figures \ref{fig:Curvature_force}a-\ref{fig:Curvature_force}d) and a group of four symmetric states ($\redtriangleleft$/$\redtriangleright$/$\bluetriangleleft$/$\bluetriangleright$: Figures  \ref{fig:Curvature_force}e-\ref{fig:Curvature_force}l), corresponding to the solutions in Figure \ref{fig:SerpentineNc1Aniso10}a. 
The jump in $\kappa_2$ at the interfaces between straight and curved segments originates from the rest configuration of serpentine strips. 
The pair of solution $\redsquare$/$\bluesquare$ is reversible about its midpoint and is invariant under the transformation $R_1$. Applying either the transformations $R_2$ or $K$ to $\bluesquare$/$\redsquare$ results in the same configuration, which is the mirror image of the original shape about the $x-z$ plane. The group of four symmetric shapes $\redtriangleleft$/$\redtriangleright$/$\bluetriangleleft$/$\bluetriangleright$ do not exhibit self-reversibility. Starting with $\bluetriangleleft$, the application of $R_1$, $R_2$, and $K$ leads to $\redtriangleright$, $\bluetriangleright$, and $\redtriangleleft$, respectively.

\begin{figure}[h!]
	\centering
	\includegraphics[width=0.8\textwidth]{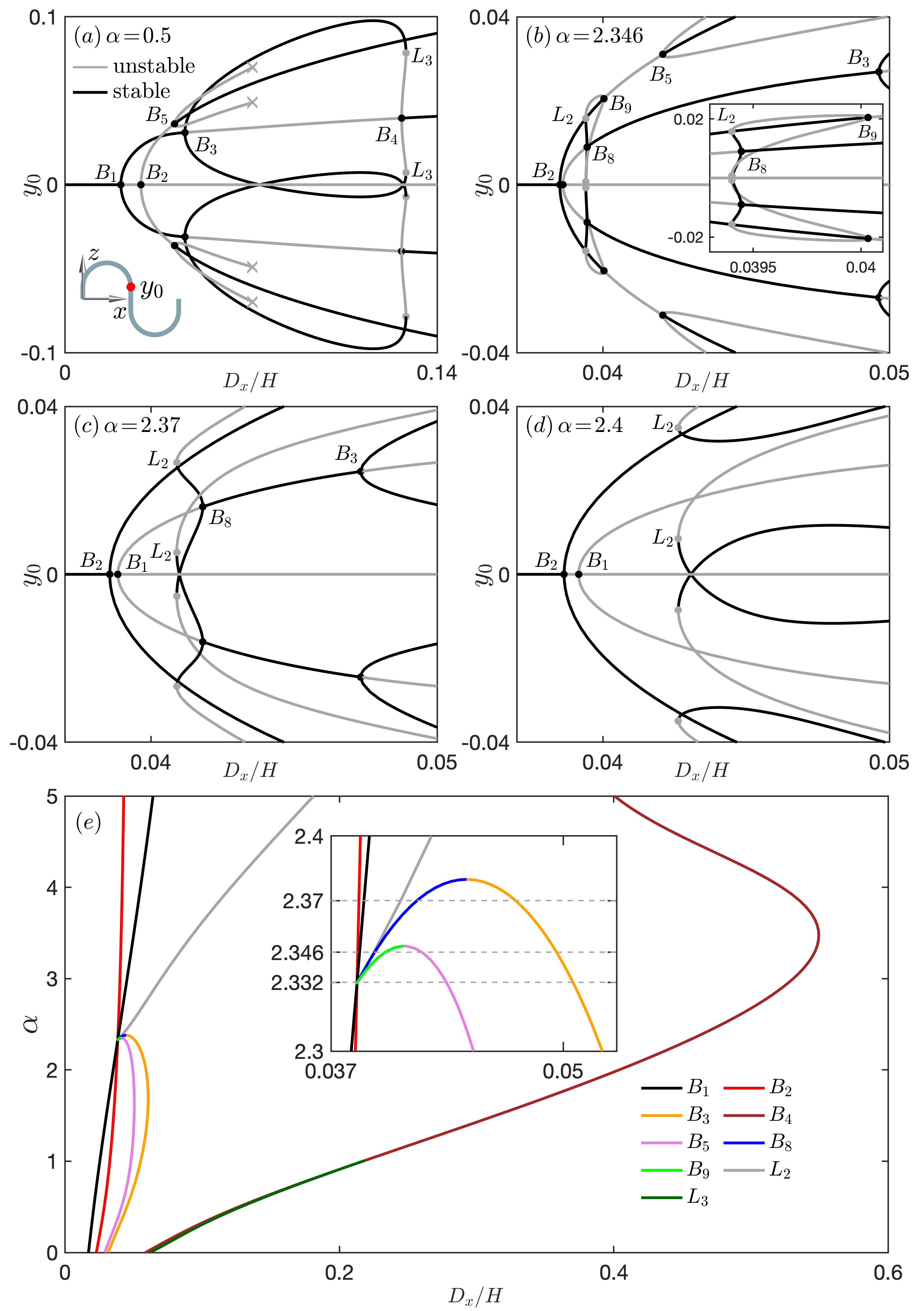}
	\caption{Detailed evolution of the buckling modes of serpentine strip with a single cell.(a-d): A series of bifurcation diagrams for different values of $\alpha$. (e) Loci of the bifurcations and folds in (a-d).} \label{fig:Nc1Aniso10phase}
\end{figure}

Figure \ref{fig:SerpentineNc1Aniso10} demonstrates that the change in buckling mode from $S$ to $M$ is caused by the switch of bifurcations $B_1$ and $B_2$. To examine this process more closely, Figure \ref{fig:Nc1Aniso10phase} presents detailed bifurcation diagrams focusing on the exchange of the first two buckling modes and displays the loci of various bifurcations/folds in the $\alpha$ versus $D_x/H$ plane. The loci curves in Figure \ref{fig:Nc1Aniso10phase}e are obtained by conducting two-parameter continuation with AUTO 07P \cite{doedel2007auto}. We find a critical $\alpha = 2.332$ where $B_1$ and $B_2$ coincide, corresponding to the intersection of the loci of $B_1$ and $B_2$ shown in Figure \ref{fig:Nc1Aniso10phase}e. For $\alpha <2.332$, $B_1$ occurs before $B_2$, and For $\alpha >2.332$, $B_2$ appears before $B_1$.   
This agrees well with the critical value of $2.4$ found in the original study by conducting finite element modeling with an increment of 0.1 for $\alpha$ \cite{zhang2013buckling}. 
At $\alpha=2.332$, the bifurcation is degenerate, and with a slight increase of $\alpha$ to 2.346 (Figure \ref{fig:Nc1Aniso10phase}b), two pairs of secondary bifurcations $B_8$ and $B_9$ emerge from the degenerate point and move onto the branches bifurcated from $B_1$ and $B_2$, respectively. Figure \ref{fig:Nc1Aniso10phase}b only labels $B_8$ and $B_9$ in the $y_0>0$ region, and there are two more identical bifurcations located symmetrically in the $y_0<0$ region. A total of eight branches bifurcate from the two pairs of $B_8$ and $B_9$, forming a closed loop. This loop shrinks as $\alpha$ decreases and disappears when $\alpha \le 2.332$.

The above evolution of buckling modes corresponds to a typical ``double eigenvalue" bifurcation, which is common in structural mechanics \cite{Bauer75multiple,Buckling88Tavener,Keener79Secondary}. A double eigenvalue bifurcation point may split into two primary bifurcation points and several secondary bifurcation points as a bifurcation parameter varies from the value at which the primary bifurcations coalesce \cite{Bauer75multiple}. Specifically, in serpentine strips, $B_1$ and $B_2$ represent the primary bifurcations, $B_8$ and $B_9$ correspond to secondary bifurcations, and $\alpha$ is the bifurcation parameter that causes the exchange of the order of the buckling modes.

With a further increase of $\alpha$ to 2.37 (Figure \ref{fig:Nc1Aniso10phase}c), the secondary bifurcation $B_9$ has coalesced with $B_5$, followed by the annihilation of $B_8$ and $B_3$ in Figure \ref{fig:Nc1Aniso10phase}d for $\alpha=2.4$. The associated secondary bifurcations $B_8$ and $B_9$ play a crucial role in achieving a smooth transition in the stability of the two buckling modes. Figure \ref{fig:Nc1Aniso10phase}e displays the loci of bifurcations and folds from Figures \ref{fig:Nc1Aniso10phase}a-\ref{fig:Nc1Aniso10phase}d. The inset illustrates the emergence of $B_8$ and $B_9$ and their subsequent annihilation with $B_3$ and $B_5$, respectively. The gray dashed lines in the inset are included solely to indicate various values of $\alpha$ and do not represent the loci of any critical points.

\begin{figure}[h!]
	\centering
	\includegraphics[width=0.9\textwidth]{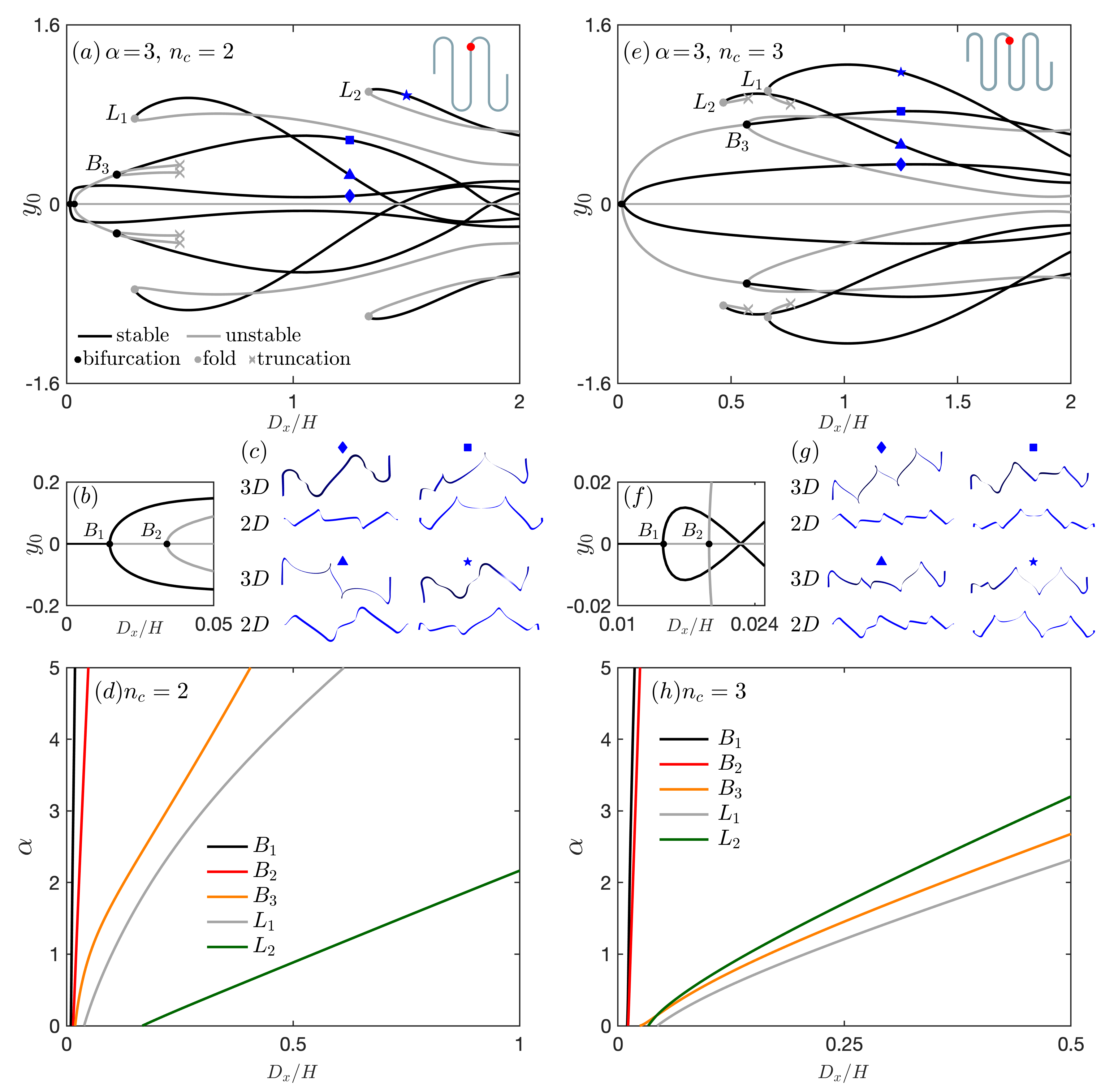}
    	\caption{Numerical results of serpentine strips for $n_c=2$ (a-d) and $n_c=3$ (e-h). $y_0$ represents the displacement of the red point in the $y$ direction, marked in the top-right inset. (a) Bifurcation diagrams with $(n_c,\alpha)=(2,3)$. (b) Inset of (a). (c) 3D renderings and their $x-y$ projections of stable configurations in (a). (d) Loci of the critical points in (a) and (b). (e) Bifurcation diagrams for $(n_c,\alpha)=(3,3)$. (f) Inset of (e). (g) 3D renderings and their $x-y$ projections of stable configurations in (e). (h) Loci of the critical points in (e) and (f).} \label{fig:Nc23Aniso10}
\end{figure}

Figure \ref{fig:Nc23Aniso10} presents numerical results for serpentine strips with $n_c \!=\! 2$ and $n_c \!=\! 3$. Figures \ref{fig:Nc23Aniso10}a-\ref{fig:Nc23Aniso10}d present the bifurcation diagram, an inset of the bifurcation diagram, renderings of several stable equilibria, and the loci of various bifurcations/folds for $n_c=2$, respectively. Figures \ref{fig:Nc23Aniso10}e-\ref{fig:Nc23Aniso10}h present similar results for $n_c=3$.  
In both scenarios, the serpentine strip first buckles into the $S$ mode ($\bluediamond$) through a supercritical pitchfork $B_1$, followed by a second bifurcation $B_2$ that creates a pair of unstable shapes. Following the unstable branches bifurcated from $B_2$, another subcritical bifurcation $B_3$ stabilizes the branch, resulting in the stable $M$ mode ($\bluesquare$).   
Figures \ref{fig:Nc23Aniso10}d and \ref{fig:Nc23Aniso10}h show that, for different values of $\alpha$, $B_1$ always occurs before $B_2$, and the loci curve never intersect within the range $\alpha \in [0 \,\, 5]$. This observation is consistent with the original study, which concluded that a serpentine strip always buckles into the $S$ mode for $n_c \geq 2$ \cite{zhang2013buckling}. Additionally, several more stable branches that were not reported in the original study \cite{zhang2013buckling} are included. Figure \ref{fig:Nc23Aniso10}a includes a pair of symmetric shape $\bluetriangle$ and a group of four symmetric shapes $\bluestar$, though only two of these branches are shown for clarity. Each of the $\bluetriangle$ and $\bluestar$ branches in Figure \ref{fig:Nc23Aniso10}e represents one state in a group of four symmetric shapes. For each case, only two branches that are symmetric about $y_0=0$ are included.

\section{tunable buckling behaviors} \label{se:tunablebuckling}

Serpentine strips tend to buckle out of plane due to their significantly higher in-plane bending stiffness compared to their out-of-plane bending stiffness and the twisting stiffness. In this section, we demonstrate that the buckling pattern and sequence of different modes in serpentine strips can be tuned through careful design of their thickness and height.

\begin{figure}[h!]
	\centering

\includegraphics[width=0.8\textwidth]{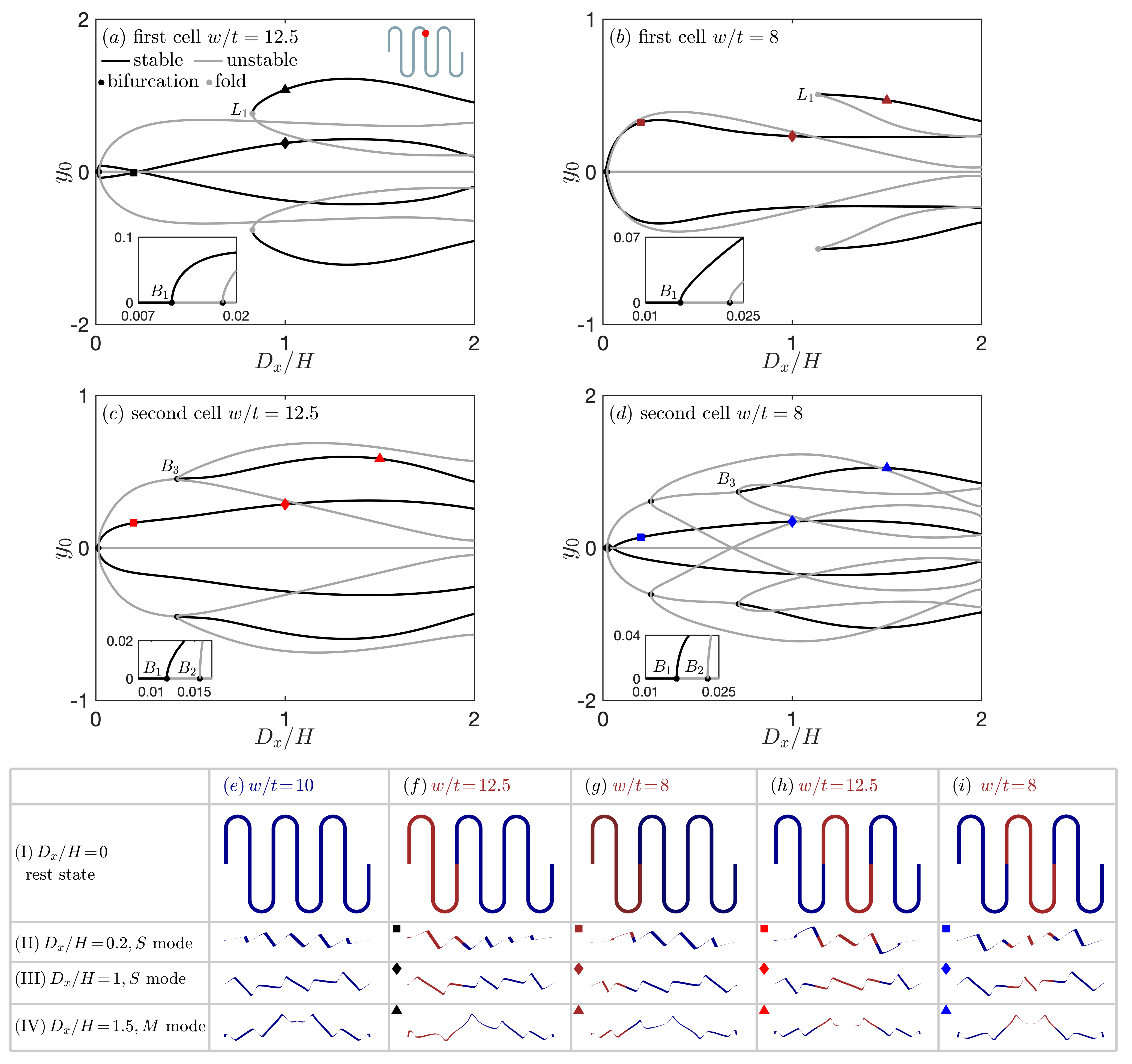}
	\caption{The buckling pattern of a serpentine strip with $n_c=3$ can be tuned through modulation of the strip thickness.(a-d) Bifurcation diagrams where the thickness of the first and second cells of the serpentine strip are varied. (e) Renderings of serpentine strips with uniform thickness. (f)-(i) Renderings corresponding to the solutions marked in (a)-(d).} \label{fig:Nc3tune}
 
\end{figure}

Figure \ref{fig:Nc3tune} illustrates the $S$ and $M$ buckling modes of a serpentine strip with three cells of varying thicknesses. The width of the strip is fixed, while its thickness is varied. The first row of the table in Figure \ref{fig:Nc3tune} summarizes the thickness distribution of five cases. In the first case, the strip has a uniform thickness with a width-to-thickness ratio ($w/t$) of 10, while in each of the subsequent cases, one cell has a different thickness. The blue cells have a default aspect ratio of $w/t=10$, and the thickness of the red cells is varied. 
Figures \ref{fig:Nc3tune}a-\ref{fig:Nc3tune}b present bifurcation diagrams for serpentine strips where the aspect ratio of the cross section of the first cell is adjusted to $w/t=12.5$ and 8, respectively. The renderings in the table represent stable configurations under different stretches. Compared to the uniform thickness case shown in Figure \ref{fig:Nc3tune}e, the deformation of the cell with a larger thickness is suppressed (as seen in Figure \ref{fig:Nc3tune}g), while the deformation of the cell with a smaller thickness is enlarged (as seen in Figure \ref{fig:Nc3tune}f). The bifurcation diagrams indicate that the structure buckles into the $S$ mode ($\blasquare$/$\bladiamond$ and $\brownsquare$/$\browndiamond$) through $B_1$, and the $M$ mode ($\blatriangle$ and $\browntriangle$) is disconnected from the planar branch, with its stability bounded by a fold point $L_1$. The self-reversibility of the thickness distribution in the rest configuration is disrupted, resulting in the buckled configuration manifesting solely as a symmetric pair about the $x-z$ plane, rather than as a group of four symmetric shapes. 

Figures \ref{fig:Nc3tune}c and \ref{fig:Nc3tune}d display bifurcation diagrams where the aspect ratio of the cross section of the second cell is adjusted to $w/t=12.5$ and 8, respectively. In these scenarios, the geometry and thickness distribution of the rest configuration retains self-reversibility about its midpoint, and the buckling behavior is similar to the case with uniform thickness. The equilibrium occurs either as a pair of symmetric shapes with self-reversibility or as a group of four symmetric shapes without self-reversibility. The structure first buckles into a stable $S$ mode through $B_1$. Shortly after $B_1$, $B_2$ emerges and bifurcates into a pair of unstable branches, which is stabilized by $B_3$, connecting to the stable $M$ mode.
With a thinner second cell, the deformation of the second cell is enlarged (Figures \ref{fig:Nc3tune}h), and vice versa (Figures \ref{fig:Nc3tune}i).

\begin{figure}[h!]
	\centering

\includegraphics[width=0.7\textwidth]{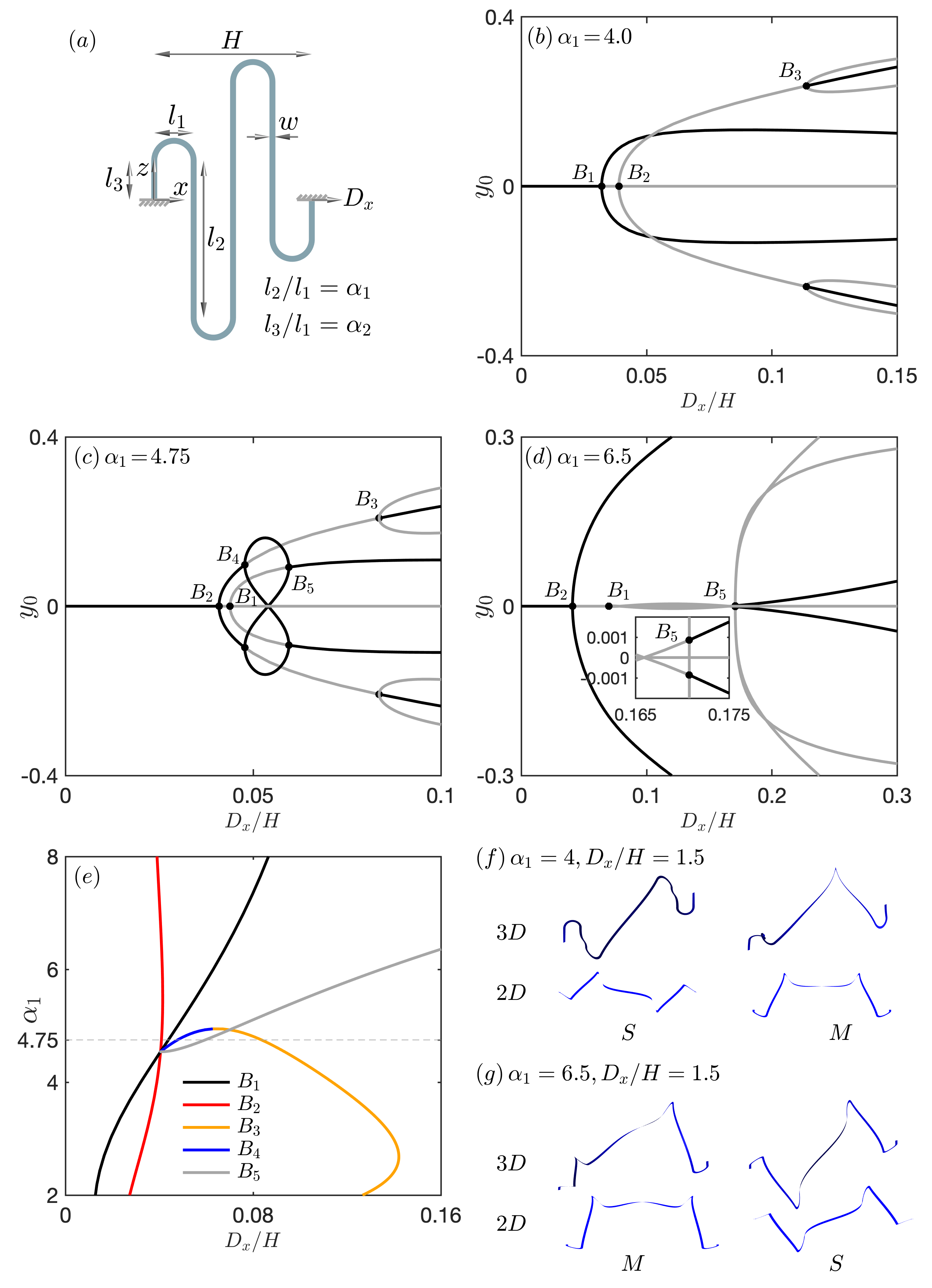}
	\caption{The buckling order of $S$ and $M$ mode in a serpentine strip with $n_c=2$ can be tuned by varying the dimensionless height $l_2/l_1$ ($l_3/l_1$ is fixed to unity).(a-d) Bifurcation diagrams with different values of $\alpha_1$. (e) Loci of bifurcation points in (a)-(d). (f) Renderings of stable states with ($\alpha_1,D_x/H=(4,1.5)$). (g) Renderings of stable states with ($\alpha_1,D_x/H=(6.5,1.5)$) } \label{fig:Nc2tuneBucklingorder}
 
\end{figure}

The original study found that the $S$ and $M$ buckling modes could exchange their order of appearance only in serpentine strips with a single cell \cite{zhang2013buckling}. In the next example, we demonstrate that the sequence of buckling modes can also be tuned for serpentine strips with two cells through the modulation of the dimensionless height. 
Figure \ref{fig:Nc2tuneBucklingorder}a displays the geometry of a serpentine strip characterized by two dimensionless heights, $\alpha_2=l_3/l_1$ and $\alpha_1=l_2/l_1$, with $\alpha_2$ fixed at unity. The antisymmetry of the structure about its midpoint is maintained. Figure \ref{fig:Nc2tuneBucklingorder}b-\ref{fig:Nc2tuneBucklingorder}d presents bifurcation diagrams with different values of $\alpha_1$. With $\alpha_1=4$, the structure first buckles into a pair of $S$ modes through a supercritical pitchfork $B_1$, and the buckled pair remains stable up to a large stretch. Following $B_1$, $B_2$ emerges and bifurcates into a pair of unstable modes, which are soon stabilized through a subcritical pitchfork $B_3$, connecting to the $M$ mode that remains stable up to a large stretch. With $\alpha_1=4.75$, $B_1$ and $B_2$ has exchanged their order of appearance, and the structure first buckles into a pair of stable $M$ mode through the supercritical pitchfork bifurcation $B_2$. The stability of the $M$ branch is soon lost through a supercritical pitchfork $B_4$ and restored through the subcritical pitchfork $B_3$. Following $B_2$, $B_1$ bifurcates into a pair of unstable modes, which are soon stabilized by a supercritical pitchfork bifurcation $B_5$, connecting to the stable $S$ mode. Similar to the case in a single-cell serpentine strip show in Figure \ref{fig:SerpentineNc1Aniso10}, the bifurcated branches from secondary bifurcations $B_4$ and $B_5$ form a closed loop. With an increase in $\alpha_1$ to 6.5, $B_3$ and $B_4$ annihilate each other, and the entire $M$ branch remains stable up to a large stretch. The loci of various bifurcations, shown in Figure \ref{fig:Nc2tuneBucklingorder}e, confirm that the exchange of the buckling modes is caused by a double eigenvalue bifurcation, which occurs at $\alpha_1=4.541$. Figures \ref{fig:Nc2tuneBucklingorder}f and \ref{fig:Nc2tuneBucklingorder}g display renderings of stable $S$ and $M$ modes for $\alpha_1=4$ and $\alpha_1=6.5$, respectively. There could be other methods to control the buckling order of $S$ and $M$ modes for serpentine strips with more than two cells. This study does not aim to conduct a comprehensive analysis but primarily focuses on illustrating the exchanging mechanism and demonstrating that it is possible to switch the sequence of $S$ and $M$ buckling modes in serpentine strips with more than one cell.

\section{Conclusions and further discussions} \label{se:discussion}

We have investigated the buckling and multistable behaviors of serpentine strips, which consist of straight and semi-circular segments, subjected to stretching. This is accomplished through a detailed comparison between experiments and numerical continuation of an anisotropic rod model. Through tabletop experiments, we have identified multiple stable states in serpentine strips by manually deforming the structures to achieve various configurations that exhibit mirror and reversible symmetries in equilibrium. These symmetries cause the equilibria in serpentine strips to occur either as pairs of states that are reversible about their midpoint or as groups of four symmetric shapes. Additionally, we observe that 
serpentine strips tend to exhibit more stable states as the stretch increases. The number of stable equilibria increased rapidly with the number of cells. Using the anisotropic rod model, we formulate the mechanics of serpentine strips as a nonlinear boundary value problem, which is solved through numerical continuation with AUTO 07P \cite{doedel2007auto}. The stability of the numerical solutions is assessed by testing for the existence of conjugate points along the length of the strip. The stable states predicted by anisotropic rod model match well with our experiments.

The results from numerical continuation further reveal that when subjected to stretch, serpentine strips undergo a series of bifurcations, each bifurcating out of plane and resulting in a pair of states. Typically, the first bifurcation is a supercritical pitchfork, creating a pair of stable states. The second bifurcation generates a pair of unstable states, which are soon stabilized through another bifurcation. These two buckling modes, arising from the first two primary bifurcations, correspond to the antisymmetric and symmetric buckling modes identified in the original study \cite{zhang2013buckling}. Our findings indicate that with the variation of the dimensionless height, the first two primary bifurcations in serpentine strips with a single cell can exchange their order of appearance. This exchange is facilitated by secondary bifurcations and corresponds to a double eigenvalue bifurcation. This process elucidates how the first two buckling modes exchange their order of appearance, a phenomenon first identified in \cite{zhang2013buckling} but not fully explained.

We further proposed controlling the buckling patterns of serpentine strips through the modulation of their thickness and height across cells. By varying the thickness of the structure, we demonstrate that the local deformation of serpentine strips with thicker or thinner cross sections could be suppressed or enlarged, respectively. 
Additionally, we find that the exchange of buckling modes could be achieved in serpentine strips with two cells by adopting a nonuniform height.  
The rich multistable behaviors and tunable buckling paths in serpentine strips could potentially inspire the design of mechanical memory devices \cite{chen2021reprogrammable,jules2022delicate,kkk2023counting}, multistable actuators \cite{chi2022bistable}, and mechanical metamaterials \cite{mei2021mechanical}.
Another promising direction for further exploration is the control of bifurcations and, consequently, the buckling pattern. 
Optimization theory and tools \cite{melot2024multi} can be employed to manage the buckling sequence and the positions of the bifurcations.

\section*{Acknowledgments}

TY used ChatGPT to improve the language and readability of the manuscript. After using this service, the authors reviewed and edited the content as needed, taking full responsibility for the content
of the publication. QS and ML acknowledge the financial support of the Shenzhen Science and Technology Innovation Commission (No. 20231115172355001). WH acknowledges the start-up funding from Newcastle University, UK.

\bibliographystyle{unsrt}
\bibliography{SerpentineStrip}

\newpage

\pagebreak
\widetext
\begin{center}
	\textbf{\large Supplemental Information} \\
\end{center}
\setcounter{equation}{0}
\setcounter{figure}{0}
\setcounter{table}{0}
\setcounter{page}{1}
\setcounter{section}{0}
\makeatletter
\renewcommand{\theequation}{S\arabic{equation}}
\renewcommand{\thefigure}{S\arabic{figure}}
\renewcommand{\bibnumfmt}[1]{[#1]}
\renewcommand{\citenumfont}[1]{#1}

\section{Formulation of serpentine strips as a two-point boundary value problem}	\label{appse:MPbvpToTPBVP} 

Equation \eqref{eq:F&Mequilibrium} represents the equilibrium condition for the balance of forces and moments on the strip. Additionally, we use unit quaternions $(q_1,q_2,q_3,q_4)$ to describe the rotation of the material frames $(\bm{d}_1,\bm{d}_2,\bm{d}_3)$ attached to the centerline of the strip. The complete set of governing equations can be written as   

\begin{equation}\label{eq:normF&M} 
\begin{aligned}
& N_1'  = N_2 \tau -N_3 \kappa_2 \, , N_2'  = -N_1 \tau + N_3 \kappa_1  \, , N_3'  =  -N_2 \kappa_1 + N_1 \kappa_2 \, , \\
& a \kappa_1' = b (\kappa_2 - \kappa_{20}) \tau - \tau \kappa_2 + N_2 \, , \\
& b \kappa_2 ' = -a \kappa_1  \tau + \tau \kappa_1  - N_1 \, , \\
& \tau'  = -b (\kappa_2 - \kappa_{20} )  \kappa_1 + a \kappa_1 \kappa_2   \,, \\
& q'_1 = \tfrac{1}{2}(-q_2 \kappa_1 -q_3 \kappa_2 -q_4 \tau)+ \mu q_1  \,,  q'_2= \tfrac{1}{2}(q_1 \kappa_1 - q_4 \kappa_2 +q_3 \tau) + \mu q_2  \,, \\
& q'_3 = \tfrac{1}{2}(q_4 \kappa_1 + q_1 \kappa_2 - q_2 \tau) + \mu q_3  \,, q'_4= \tfrac{1}{2}(-q_3 \kappa_1 +q_2 \kappa_2 +q_1 \tau) + \mu q_4  \,, \\
& x' =2( q_2 q_4 + q_1 q_3) \,, \; y' =2( q_3 q_4 - q_1 q_2 ) \,,\; z ' =2( q_1^2 + q_4^2-\tfrac{1}{2} ) \,, \mu'=0 \,. \\
\end{aligned}
\end{equation}

Following Healey and Mehta \cite{healey2006straightforward}, the dummy parameter $\mu$ in Equation \eqref{eq:normF&M} allows for a consistent prescription of boundary conditions for quaternions. We introduce a trivial equation $\mu'=0$ and treat $\mu$ as a scalar unknown. During numerical continuation, we monitor $\mu$ to ensure its value remains numerically zero \cite{healey2006straightforward}. 
For rods with a rectangular cross section composed of an elastically isotropic material, the bending and twisting stiffness are given by \cite{timoshenko1951theory},

\begin{equation}\label{eq:inertiaofmoment} 
\begin{aligned} 
EI_1=\tfrac{1}{12} E w t^3\,,\;  EI_2=\tfrac{1}{12} E w^3 t \,,\; 
GJ=\lambda G w t^3=\lambda w \frac{E}{2(1+\nu)} t^3 \, ,
\end{aligned}
\end{equation}
which leads to 
\begin{equation}\label{eq:coefficients} 
a=\frac{EI_1}{GJ}=\frac{(1+\nu)}{6\lambda} \,,\;  b=\frac{EI_2}{GJ}=\frac{(1+\nu)}{6\lambda}\left(\frac{w}{t} \right)^2 \, .
\end{equation}

Here, $\nu$ is the Poisson's ratio, which is set to 0.33 in this study. $\lambda$ corresponds to a shape factor and depends on the aspect ratio of the cross section $w/t$ through the following relationship \cite{timoshenko1951theory},

\begin{equation}\label{eq:shapefactor} 
\begin{aligned} 
\lambda=\tfrac{1}{3} \left( 1-\tfrac{192}{\pi^5} \tfrac{t}{w} \sum _{k=1} ^{\infty} \tfrac{1}{(2k-1)^5} \tanh \left( \tfrac{\pi (2k-1) w}{2t} \right)  \right) \, .
\end{aligned}
\end{equation}

To address the discontinuity of the in-plane bending curvature of serpentine strips at the interfaces between straight and circular segments, we divide each serpentine cell into five segments and assign a full set of governing equations \eqref{eq:normF&M} to each segment. The governing equation of the $i_{\text{th}}$ segment can be rewritten as follows, 

\begin{equation}\label{eq:twopointBVP} 
\begin{aligned}
& N_{1i}'  = ( N_{2i} \tau_i -N_{3i} \kappa_{2i} )u_i \, , N_{2i}'  =( -N_{1i} \tau_i + N_{3i} \kappa_{1i} )u_i \, , N_{3i}'  = ( -N_{2i} \kappa_{1i} + N_{1i} \kappa_{2i} ) u_i\, , \\
& a \kappa_{1i}' =( b (\kappa_{2i} - \kappa_{20}) \tau_i - \tau_i \kappa_{2i} + N_{2i} ) u_i\, , \\
& b (\kappa_{2i} ' - \kappa_{20} ')  =( -a \kappa_{1i}  \tau_i + \tau_i \kappa_{1i}  - N_{1i} ) u_i\, , \\
& \tau_i'  =( -b (\kappa_{2i} - \kappa_{20} )  \kappa_{1i} + a \kappa_{1i} \kappa_{2i} ) u_i \,, \\
& q'_{1i} = \tfrac{1}{2}(-q_{2i} \kappa_{1i} -q_{3i} \kappa_{2i} -q_{4i} \tau_i) u_i+ \mu q_{1i} u_i  \,,  q'_{2i}= \tfrac{1}{2}(q_{1i} \kappa_{1i} - q_{4i} \kappa_{2i} +q_{3i} \tau_i) u_i + \mu q_{2i} u_i  \,, \\
& q'_{3i} = \tfrac{1}{2}(q_{4i} \kappa_{1i} + q_{1i} \kappa_{2i} - q_{2i} \tau_i) u_i+ \mu q_{3i} u_i  \,, q'_{4i}= \tfrac{1}{2}(-q_{3i} \kappa_{1i} +q_{2i} \kappa_{2i} +q_{1i} \tau_i) u_i+ \mu q_{4i} u_i  \,, \\
& x_i' =2( q_{2i} q_{4i} + q_{1i} q_{3i}) u_i\,, \; y_i' =2( q_{3i} q_{4i} - q_{1i} q_{2i} ) u_i \,,\; z_i ' =2( q_{1i}^2 + q_{4i}^2-\tfrac{1}{2} ) u_i \,, \mu'=0 \,. \\
\end{aligned}
\end{equation}
where a prime denotes a derivative with respect to $\bar{s}$ ($\bar{s} \in [0,1]$ ), $u_i$ corresponds to length of the $i_{\text{th}}$ segment, and $i \in [1,5 n_c]$. In numerical continuation, we set the length of the semi-circle to unity. The introduction of the scaling factor $u_i$ normalizes the integral interval of all segments to $[0 \, 1]$. We impose clamping boundary conditions at the two ends of serpentine strips and specify continuity conditions for forces, moments, positions, and material frames at the internal boundaries corresponding to straight-circular interfaces. Essentially, all boundary and continuity conditions are imposed at the $``0"$ and $``1"$ ends, resulting in a standard two-point boundary value problem. We use the rest configuration of the serpentine strips as the initial solution for numerical continuation. Details of the specification of boundary conditions and initial solutions are omitted here.

\section{Stability test}	\label{appse:stabilitytest} 

To determine the stability of the equilibria obtained through numerical continuation, we adopt a method from geometric mechanics to test for the existence of a conjugate point \cite{borum2020helix,borum2020infinitely}. The stability test we implemented is applicable to a strip with clamping boundary conditions. 
Equation \eqref{eq:normF&M} can be rewritten as follows,
\begin{equation}\label{eq:normF&M2} 
\begin{aligned}
M_3' = M_1 (\frac{M_2}{b} +\kappa_{20}) - M_2  \frac{M_1}{a}
\, ,\\
M_1 ' =M_2 \frac{M_3}{c} - M_3 (\frac{M_2}{b} + \kappa_{20}) + N_2 \, , \\
M_2 '=- M_1 \frac{M_3}{c} + M_3 \frac{M_1}{a} - N_1 \, , \\
N_3' =  N_1 (\frac{M_2}{b} +\kappa_{20}) - N_2 \frac{M_1}{a}  \, , \\
N_1' = N_2 \frac{M_3}{c} -N_3 (\frac{M_2}{b} +\kappa_{20}) \, , \\
N_2'  = N_3 \frac{M_1}{a}  - N_1 \frac{M_3}{c} \, ,\\
\end{aligned}
\end{equation}
where $c$ corresponds to the normalized torsional rigidity, i.e., $c=1$. $\kappa_{20}$ is a piecewise function representing the in-plane bending curvature of the rest configuration. The bending and twisting rigidity can also be a piecewise function, as in the case with nonuniform thickness discussed in Section \ref{se:tunablebuckling}. Specifically, to determine the stability of a solution $(\bm{M}, \bm{N})$ from Equation \eqref{eq:normF&M}, we solve the following matrix differential equations,

\begin{equation}\label{appeq:IVP} 
\begin{aligned}
D'=FD \,, P'=QD + HP \,,\\
\end{aligned}
\end{equation}
where a prime denotes a derivative with respect to $s$ ($s \in [0 \,, L]$). The coefficient matrices $F$, $Q$, and $H$ can be written as \cite{borum2020helix,borum2020infinitely}

\begin{equation}\label{appeq:F2} 
\begin{aligned}
F
&=
\begin{bmatrix} 0 & M_2(\frac{1}{b} - \frac{1}{a}) + \kappa_{20} & M_1 (\frac{1}{b} - \frac{1}{a}) & 0 & 0 & 0 \vspace{4pt} \\
M_2 (\frac{1}{c} - \frac{1}{b}) - \kappa_{20} & 0 & M_3 (\frac{1}{c} - \frac{1}{b}) & 0 & 0 & 1 \vspace{4pt} \\ 
M_1 (\frac{1}{a} - \frac{1}{c}) & M_3 (\frac{1}{a} - \frac{1}{c}) & 0  & 0 & -1 & 0 \vspace{4pt} \\
0 & -\frac{N_2}{a} & \frac{N_1}{b} & 0 & \kappa_2 & - \kappa_1 \vspace{4pt} \\
\frac{N_2}{c} & 0 & -\frac{N_3}{b} & - \kappa_2 & 0 & \tau \vspace{4pt} \\
-\frac{N_1}{c} & \frac{N_3}{a} & 0 & \kappa_1 &- \tau & 0 
\end{bmatrix} 
\end{aligned}
\end{equation}

\begin{equation}\label{appeq:Q} 
\begin{aligned}
Q	
&=
\begin{bmatrix} 1/c & 0 & 0 & 0 & 0 & 0 \\
0 & 1/a & 0 & 0 & 0 & 0 \\
0 & 0 & 1/b & 0 & 0 & 0 \\
0 & 0 & 0 & 0 & 0 & 0 \\
0 & 0 & 0 & 0 & 0 & 0 \\
0 & 0 & 0 & 0 & 0 & 0 \\
\end{bmatrix} \,,
H
&=
\begin{bmatrix} 0 & \kappa_2 & -\kappa_1 & 0 & 0 & 0 \\
- \kappa_2 & 0 & \tau & 0 & 0 & 0 \\ 
\kappa_1 & -\tau & 0  & 0 & 0 & 0 \\ 
0 & 0 & 0 & 0 & \kappa_2 & -\kappa_1 \\
0 & 0 & 1 & -\kappa_2 & 0 & \tau \\
0 & -1 & 0 & \kappa_1 & -\tau & 0 
\end{bmatrix} \,,
\end{aligned}
\end{equation}		
where $\kappa_1= \frac{M_1}{a} $, $\kappa_2= \frac{M_2}{b} + \kappa_{20}$, and $\tau= \frac{M_3}{c} $. We first solve Equation \eqref{eq:normF&M} subject to boundary conditions by conducting numerical continuation with AUTO 07P \cite{doedel2007auto}. Then, combining the initial conditions $D(0)=I_{6 \times 6}$ and $P(0)=0_{6 \times 6}$, along with $(\bm{M}(0),\bm{N}(0))$, we solve the initial value problem (\ref{eq:normF&M2}-\ref{appeq:IVP}) using Matlab ODE45.
If the solution satisfies det(P)$\ne 0$ for all $s \in (0,L]$, then the equilibrium is stable. If det(P)$=0$ for some $s \in (0,L]$, then the equilibrium is unstable.

\begin{figure}[h!]
	\centering
	\includegraphics[width=0.8\textwidth]{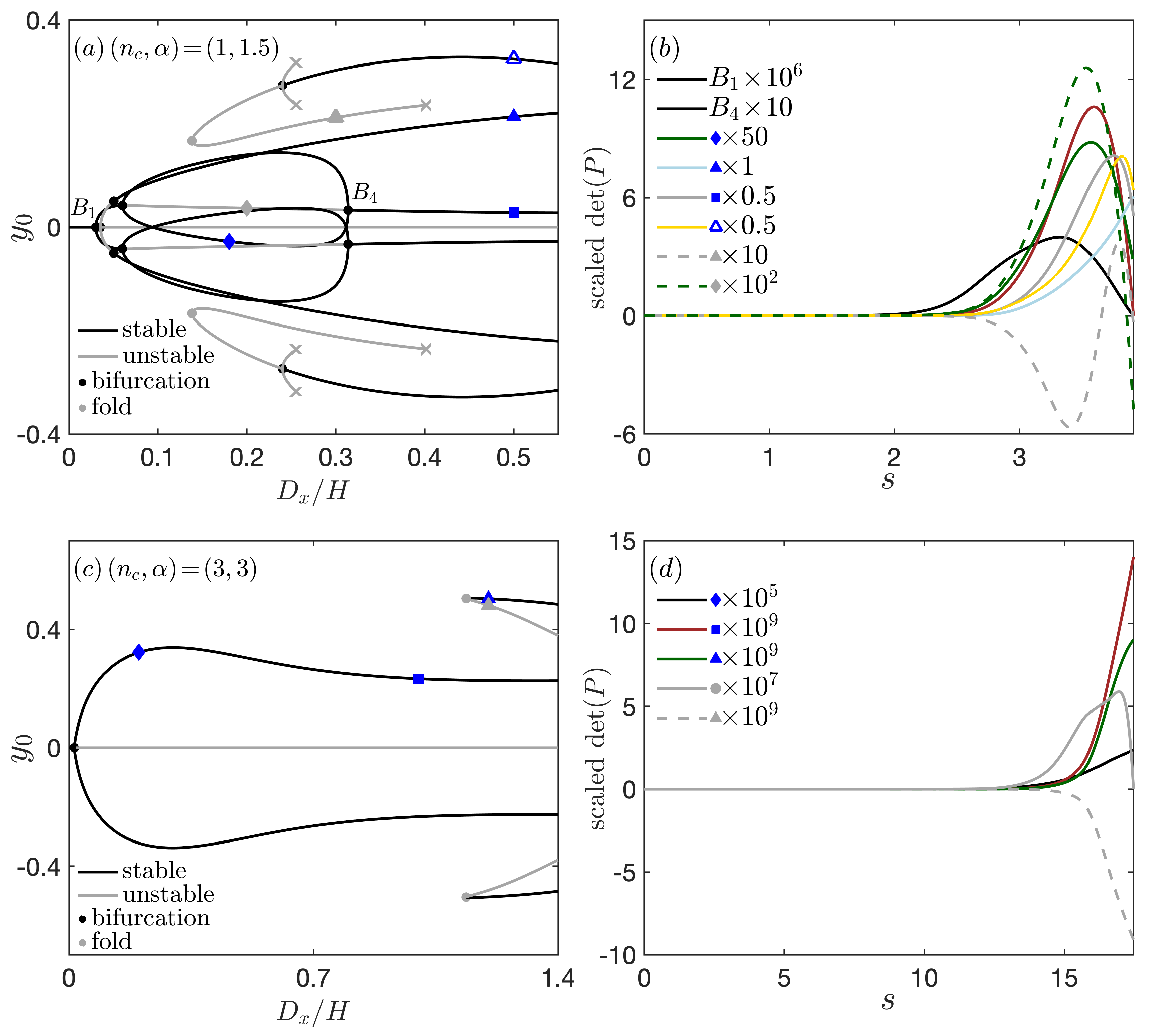}
	\caption{Stability test of equilibria of serpentine strips.  (a) Bifurcation curves from Figure \ref{fig:SerpentineNc1Aniso10}a.(b) Conjugate point test of equilibria in (a). (c) Bifurcation curves from Figure \ref{fig:Nc3tune}b. (d) Conjugate point test of equilibria in (c). } \label{fig:Nc1Nc3stability}
\end{figure}

Here we present several examples that apply the conjugate point test to determine the stability of equilibria obtained from AUTO (see Figure \ref{fig:Nc1Nc3stability}). Figures \ref{fig:Nc1Nc3stability}a and \ref{fig:Nc1Nc3stability}c display parts of the bifurcation diagrams from Figures \ref{fig:SerpentineNc1Aniso10}a and \ref{fig:Nc3tune}b, respectively. Figures \ref{fig:Nc1Nc3stability}b and \ref{fig:Nc1Nc3stability}d present the conjugate point test results for stable and unstable solutions from Figures \ref{fig:Nc1Nc3stability}a and \ref{fig:Nc1Nc3stability}c, respectively. For critical points like bifurcations and folds, the stability test curve crosses zero right at the end $s=L$. All stable solutions do not cross zero in the interval $[0 \,, L]$.

\end{document}